\documentclass[trackchanges]{aastex701}

\usepackage[whole]{bxcjkjatype}

\begin{document}

\title{ALMA Band 9 CO(6--5) Reveals a Warm Ring Structure Associated with the Embedded Protostar in the Cold Dense Core MC~27/L1521F}

\author[orcid=0000-0002-2062-1600,sname='Kazuki Tokuda']{Kazuki Tokuda}
\affiliation{Faculty of Education, Kagawa University, Saiwai-cho 1-1, Takamatsu, Kagawa 760-8522, Japan}
\email[show]{tokuda.kazuki@kagawa-u.ac.jp}

\author[0000-0002-7951-1641]{Mitsuki Omura}
\affiliation{Department of Earth and Planetary Sciences, Faculty of Science, Kyushu University, Nishi-ku, Fukuoka 819-0395, Japan}
\email[hide]{}

\author[0000-0002-8217-7509]{Naoto Harada}
\affiliation{Department of Astronomy, Graduate School of Science, The University of Tokyo, 7-3-1 Hongo, Bunkyo-ku, Tokyo 113-0033, Japan}
\email[hide]{}

\author[0000-0001-6580-6038]{Ayumu Shoshi}
\affiliation{Department of Earth and Planetary Sciences, Faculty of Science, Kyushu University, Nishi-ku, Fukuoka 819-0395, Japan}
\email[hide]{}

\author[0009-0005-4458-2908]{Naofumi Fukaya}
\affiliation{Department of Physics, Nagoya University, Furo-cho, Chikusa-ku, Nagoya 464-8601, Japan}
\email[hide]{}

\author[0000-0001-7826-3837]{Toshikazu Onishi}
\affiliation{Department of Physics, Graduate School of Science, Osaka Metropolitan University, 1-1 Gakuen-cho, Naka-ku, Sakai, Osaka 599-8531, Japan}
\email[hide]{}

\author[0000-0002-1411-5410]{Kengo Tachihara}
\affiliation{Department of Physics, Nagoya University, Furo-cho, Chikusa-ku, Nagoya 464-8601, Japan}
\email[hide]{}

\author[0000-0003-1549-6435]{Kazuya Saigo}
\affiliation{Graduate School of Science and Engineering, Kagoshima University, 1-21-40 Korimoto Kagoshima-city Kagoshima, 890-0065, Japan}
\email[hide]{}

\author[0000-0002-8125-4509]{Tomoaki Matsumoto}
\affiliation{Faculty of Sustainability Studies, Hosei University, Fujimi, Chiyoda-ku, Tokyo 102-8160, Japan}
\email[hide]{}

\author[0000-0002-8966-9856]{Yasuo Fukui}
\affiliation{Department of Physics, Nagoya University, Furo-cho, Chikusa-ku, Nagoya 464-8601, Japan}
\email[hide]{}

\author[0000-0001-7813-0380]{Akiko Kawamura}
\affiliation{National Astronomical Observatory of Japan, National Institutes of Natural Sciences, 2-21-1 Osawa, Mitaka, Tokyo 181-8588, Japan}
\email[hide]{}

\author[0000-0002-0963-0872]{Masahiro N. Machida}
\affiliation{Department of Earth and Planetary Sciences, Faculty of Science, Kyushu University, Nishi-ku, Fukuoka 819-0395, Japan}
\email[hide]{}

\begin{abstract}
Infall and outflows, coupled with magnetic fields, rapidly structure the gas around newborn protostars. Shocks from interacting components encode the temperature and density distribution, offering a direct probe of the earliest evolution history. However, interferometric observations characterizing warm envelopes using high-excitation lines remain scarce. We present ALMA Band~9 observations of the Taurus dense core MC~27/L1521F, which hosts a Class~0 protostar, targeting the CO($J$=6--5) line at an angular resolution of $\sim$2\arcsec\ ($\approx$300~au). We detect an off-centered ring-like structure with a diameter of $\sim$1000~au that was not identifiable in previous low-$J$ CO data, where emission close to the systemic velocity is strongly affected by optical depth. The ring shows a typical peak brightness temperature of $\sim$3~K at our resolution. Excitation considerations indicate that the detected CO($J$=6--5) emission likely arises from relatively warm ($T \gtrsim 20$~K) and dense ($n({\rm H_2}) \gtrsim 10^{5}$~cm$^{-3}$) gas embedded within the surrounding cold, dense core. The morphology and kinematics suggest an energetic and localized shock-heating event, potentially linked to dynamical gas--magnetic-field interactions in the earliest protostellar phase. Our results demonstrate that high-$J$ CO observations provide a powerful new window on warm and dense gas components, enabling a more direct view of the physical processes operating at the onset of star formation.
\end{abstract}


\section{Introduction}\label{sec:intro}

The physical conditions in dense cores immediately before and after protostar formation remain among the least constrained stages of star formation. At this earliest epoch, the onset of outflows coupled with magnetic fields can operate simultaneously, and thus the gas density and velocity fields are expected to evolve rapidly and become highly non-axisymmetric \citep[e.g.,][]{Matsumoto11,Machida_2020}. Observational constraints on the temperature and density structure in this phase are therefore essential for reconstructing the initial accretion/outflow history and the physical pathways toward disk formation. A key observational challenge is to identify the warm and dynamically processed gas \cite[e.g.,][]{Shinnaga_2009,Tokuda_2018} produced by shocks and feedback near newborn protostars. Commonly used millimeter tracers face inevitable limitations in isolating such dynamically processed gas. Low-$J$ CO lines are often optically thick and prone to self-absorption. While other dense gas tracers alleviate the optical-depth problem, they do not necessarily provide a selective view of moderately dense and heated gas components.

High-excitation CO transitions at submillimeter wavelengths provide a complementary probe because they preferentially trace warmer and/or denser gas, including components heated by shocks or UV irradiation associated with feedback. Single-dish surveys with Herschel and ground-based facilities have demonstrated that high-$J$ CO lines are powerful diagnostics of warm molecular gas in low-mass protostars and their outflow cavities (e.g., \citealt{vanKempen_2009a,vanKempen_2009b,Yildiz_2013,Yang_2018,Kang_2021}).  In contrast, interferometric observations of such high-excitation lines at higher angular resolution have remained limited \citep{Kristensen_2013,Codella_2023}, leaving the spatial origin and morphology of warm gas near the youngest protostars relatively unexplored.

MC~27 \citep{Mizuno94,Onishi99,Onishi02}, also known as L1521F \citep{Codella97}, stands out as one of the densest cores within the Taurus molecular cloud ($D \sim$140\,pc; \citealt{Galli18}). It harbors a notably low-luminosity ($\lesssim$0.07\,$L_{\odot}$) Class~0 protostar \citep{Bourke06}. Single-dish observations have consistently reported both the gas and dust temperatures around a $\sim$10 K \citep[e.g.,][]{Codella97,Tatematsu04,Kirk_2007}, suggesting that the majority of the gas and dust within this line of sight reside in a cold environment. ALMA observations revealed a protostellar source MMS--1, two starless condensations MMS--2, 3, and complex arc-like structures on $\sim$10$^3$--10$^4$ au scales, suggesting that the initial conditions of star formation can be highly dynamical in this core (\citealt{Tokuda_2014,Tokuda_2016}; see also theoretical interpretation by \citealt{Matsumoto_2015, Matsumoto_2017, Machida_2020,Machida_2025}). Subsequent high-resolution studies further established the compact protostellar disk component and emphasized the presence of localized warm CO gas potentially linked to dynamical gas interactions (\citealt{Tokuda_2017,Tokuda_2018}).  More recently, further high-resolution ALMA imaging resolved a tiny $\sim$10 au disk and spike-like structures that were interpreted in terms of episodic magnetic-flux transport events (\citealt{Tokuda_2024}), highlighting the potential importance of non-steady, magnetically mediated processes in MC~27.

Ring-like gas structures have also emerged as a potentially important signature of magnetic-flux transport in young stellar objects (e.g., the 7000~au-scale C$^{18}$O($J$=2--1) ring around IRS~2 in Corona Australis; \citealt{Tokuda_23interC}).  Motivated by these developments, we target MC~27 with ALMA Band~9 to image the CO($J$=6--5) line at $\sim$2\arcsec\ resolution ($\sim$300\,au), aiming to reveal warm molecular gas components that may be hidden or severely distorted in low-$J$ CO observations near $V_{\rm sys}$.  In this paper, we present the CO($J$=6--5) imaging results and discuss their implications for energetic, localized heating and dynamical processes potentially regulated by magnetic field operating at the onset of star formation.

\section{Observations and Data Reduction} \label{sec:data}

MC~27 was observed as part of an ALMA Cycle~11 program (Project code: 2024.1.00023.S) using the Atacama Compact Array (ACA) 7\,m array in Band~9. The observations were executed in five execution blocks between 2025 September 11 and 23. The representative sky frequency was 691.44\,GHz. 
The dataset was taken in a band-to-band phase-referencing mode. 
Four spectral windows were configured: two wide windows centered at 673.754\,GHz and 685.421\,GHz (each 1.875\,GHz bandwidth; 1920 channels), one continuum window centered at 689.364\,GHz (2.000\,GHz bandwidth; 128 channels), and one high-resolution window covering CO($J{=}6$--5) centered at 691.437\,GHz with 0.117\,GHz bandwidth and 480 channels.

We started from the pipeline-calibrated visibilities delivered by ALMA (QA2 PASS). We re-imaged the calibrated measurement sets on the user side using CASA version 6.7.2. Imaging was performed with \texttt{tclean} in interactive mode (\texttt{iclean}); the cleaning mask was selected manually, and the cleaning threshold was set to $\sim$0.5 times the rms noise measured in emission-free channels of the dirty cube. We adopted natural weighting to maximize surface-brightness sensitivity, and imaged the CO($J{=}6$--5) cube with a velocity resolution of 0.12\,km\,s$^{-1}$. The achieved rms noise in the final cube is 0.14\,K. The restoring beam is $1\farcs8 \times 1\farcs3$ at a position angle of $-6\fdg1$. For the continuum, we combined three spectral windows, yielding a total bandwidth of 5.75\,GHz with an effective central frequency of 682.5\,GHz. The continuum image reaches an rms sensitivity of 3.7\,mJy\,beam$^{-1}$, with a synthesized beam of $1\farcs7 \times 1\farcs2$ at a position angle of $-5\fdg2$.

\begin{figure}[htbp]
    \centering
    \includegraphics[width=0.8\columnwidth]{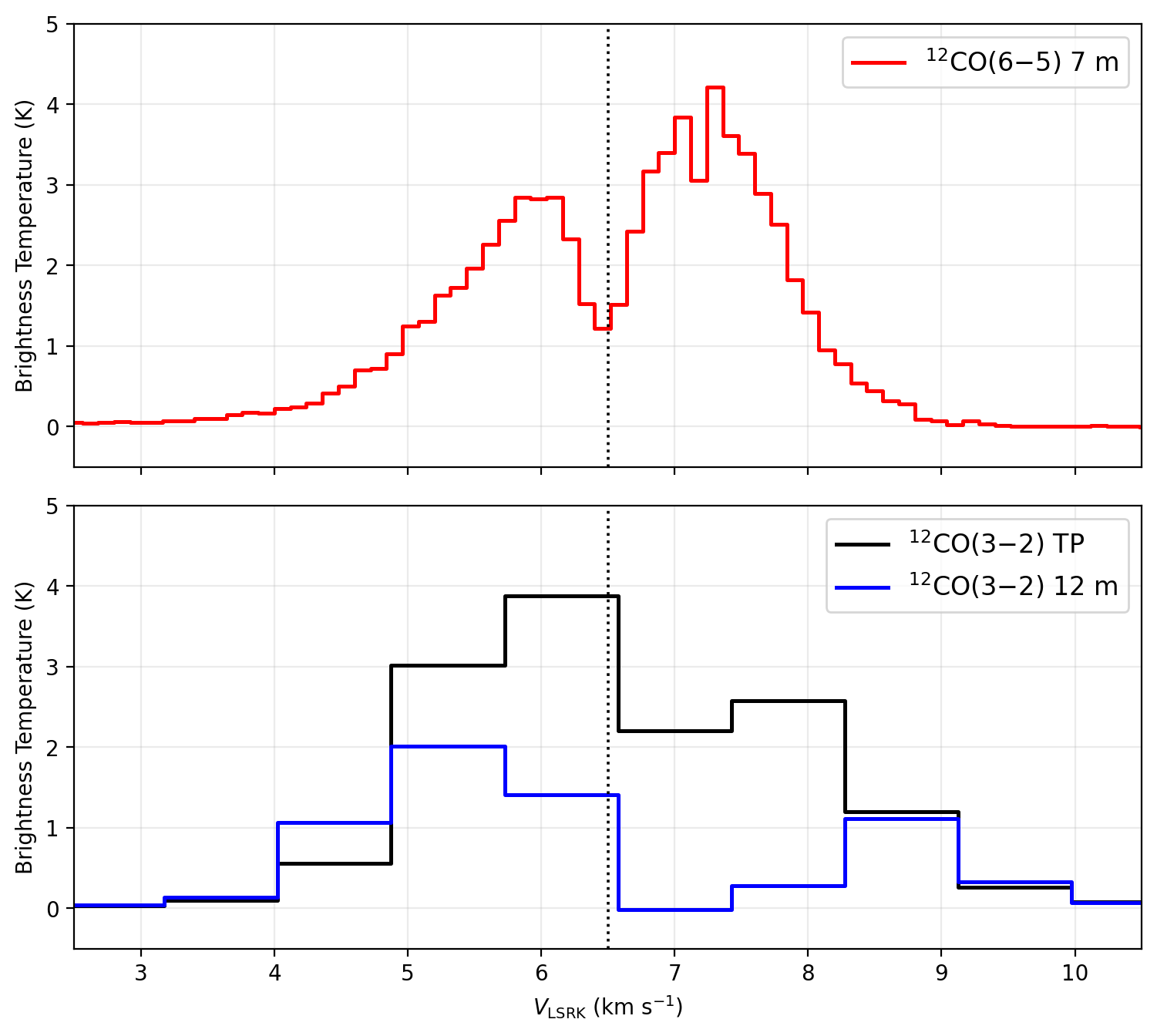}
    \caption{
     \textit{Top:} Mean CO($J$=6--5) spectra extracted from a circular aperture of radius $5\arcsec$ centered at (ICRS) $\alpha=04^{\rm h}28^{\rm m}39\fs97$, $\delta=+26^\circ51^\prime21\farcs5$.  The systemic velocity of the core, $V_{\rm sys}=6.5$~km~s$^{-1}$, is indicated by the vertical dotted line.
     \textit{Bottom:} Mean CO($J$=3--2) spectrum obtained with the previous study \cite{Tokuda_2018}, averaged within a circular aperture of radius $10\arcsec$ centered at the same position. The black histogram shows the Total Power (TP) spectrum, while the blue histogram shows the ALMA 12~m array spectrum. Both spectra are displayed at a velocity resolution of 0.85~km~s$^{-1}$.}
     \label{fig:spect}
\end{figure}

At present, ALMA does not offer Total Power (TP) Array observations in Band~9, and our ACA 7\,m array data therefore lack zero spacings. As a consequence, the map fidelity and missing flux cannot be fully quantified with high precision from the ALMA data alone. Fortunately, single-dish CO($J$=6--5) observations toward MC~27 have been reported with the Caltech Submillimeter Observatory (CSO) by \citet{Shinnaga_2009} (see their Figure~1). In the CSO $10\arcsec$ beam, the spectrum toward the protostellar position reaches a main-beam brightness temperature of $\sim$5~K with an rms of $\sim$0.6--1.4~K. For a rough consistency check, we derived an ALMA ACA spectrum averaged within a $10\arcsec$ beam ($5\arcsec$ radius); the peak brightness temperature is $\sim$4~K (see Figure~\ref{fig:spect}). The ACA spectrum also shows a depression of $\sim$2~K (in relative intensity) near the systemic velocity. While the signal-to-noise ratio of the CSO spectrum is insufficient for a detailed channel-by-channel comparison of the line profile, the overall agreement suggests that any missing flux in the ACA data is likely at most a few tens of percent when integrated over the full velocity range. In addition, compared to the previously published CO($J$=3--2) spectra, the CO($J$=6--5) spectrum clearly detects emission near the systemic velocity, $V_{\rm sys}=6.5~{\rm km~s^{-1}}$ \citep{Onishi99,Tatematsu04}, that was largely absent in interferometric observations at lower excitation, where the systemic-velocity component is strongly affected by optical depth and self-absorption. Although the relatively coarse velocity resolution of the CO($J$=3--2) data (0.85~km~s$^{-1}$) makes a strict channel-by-channel comparison difficult, the emission around $V_{\rm LSRK}\sim7~{\rm km~s^{-1}}$ is evidently filtered out in the CO($J$=3--2) interferometric (12~m array) spectrum (see Figure~\ref{fig:spect}).

\section{Results}\label{sec:Results}

\begin{figure}[htbp]
    \centering
    \includegraphics[width=1.0\columnwidth]{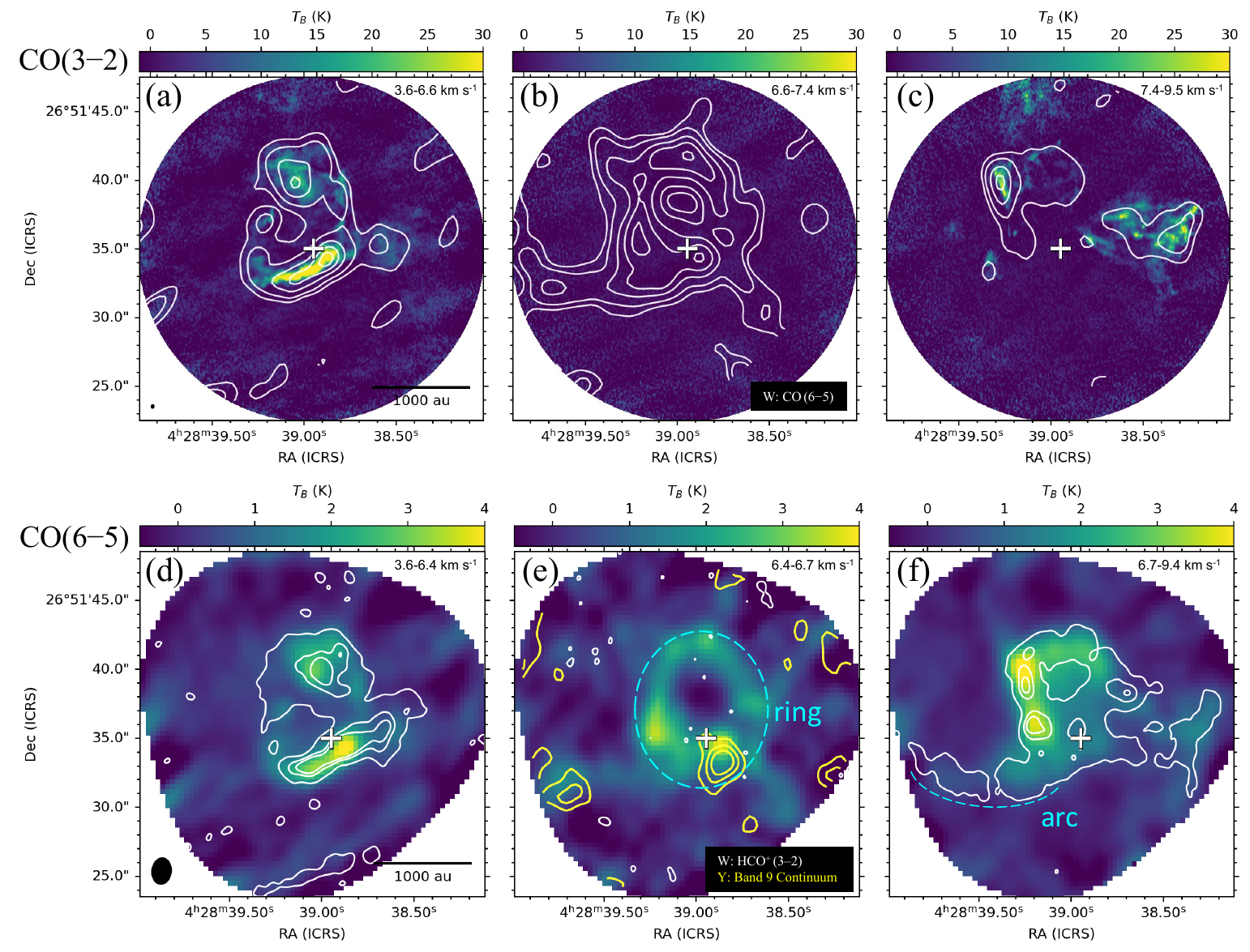}
    \caption{
    Velocity-binned intensity maps comparing the previously obtained CO($J$=3--2) and HCO$^{+}$($J$=3--2) emission with new ALMA Band~9 CO($J$=6--5) and continuum data. 
    {\bf (a--c)} CO($J$=3--2) intensity maps \citep{Tokuda_2018} shown in color, with white contours showing the CO($J$=6--5) averaged over the same velocity intervals. The velocity ranges are indicated at the top of each panel. The middle-velocity panel (b) corresponds to the range where the CO($J$=3--2) emission is strongly suppressed by self-absorption; the blueshifted and redshifted components relative to this interval are shown in panels (a) and (c), respectively. 
    The CO($J$=6--5) contour levels are [0.5, 1.0, 2.0, 3.0, 4.0]~K.
    {\bf (d--f)} CO($J$=6--5) intensity maps are shown in color. White contours show the HCO$^{+}$($J$=3--2) emission \citep{Tokuda_2014}, while yellow contours in panel (e) indicate the ALMA Band~9 continuum emission. The velocity ranges are shown at the top of each panel; the intermediate velocity interval in panel (e) is defined based on the HCO$^{+}$ data. The HCO$^{+}$ contour levels are [0.5, 2.0, 3.0]~K, and the Band~9 continuum contour levels are [11, 22, 33]\,mJy\,beam$^{-1}$.
    In all panels, the cross marks the protostellar position. The synthesized beam in the color-scale data is shown in the lower-left corner of the relevant panels. The reduced ALMA Band 9 continuum 2D image and the CO($J$=6--5) data cube used to generate the results shown here are available as Data behind the Figure in the FITS format.
    }
    \label{fig:COHCOP}
\end{figure}

We present the ALMA ACA Band~9 results of CO($J$=6--5) line emission and the associated continuum toward MC~27/L1521F. To facilitate a direct comparison with structures reported in previous studies, Figure~\ref{fig:COHCOP} illustrates our new CO($J$=6--5) maps with the CO($J$=3--2) data from \citet{Tokuda_2018} and the HCO$^{+}$($J$=3--2) data from \citet{Tokuda_2014}. As also shown in Figure~\ref{fig:spect}, the full velocity extent over which CO($J$=6--5) is detected is approximately $V_{\rm LSR}\sim3$--$9~{\rm km~s^{-1}}$, comparable to that seen in CO($J$=3--2) and HCO$^{+}$($J$=3--2). The CO($J$=6--5) emission exhibits a rich and highly structured morphology; for example, Figure~\ref{fig:COHCOP}(f) show indications of the arc-like feature extending westward from the protostar that was previously identified in HCO$^{+}$ \citep{Tokuda_2014}, which is also detected in CO($J$=6--5).

A particularly notable feature is found near the intermediate velocities: in CO($J$=3--2) (panel~b) and in HCO$^{+}$($J$=3--2) (panel~e), the corresponding channels are strongly affected by self-absorption and are largely filtered out in interferometric imaging, whereas the CO($J$=6--5) emission remains detectable over the same velocity range. 
Notably, in Figure~\ref{fig:COHCOP}(e) ($V_{\rm LSRK}=6.4-6.7~{\rm km~s^{-1}}$), the combination of the ring and the eastern arc yields a pattern that, as an aside, resembles the Arabic numeral ``9''. By contrast, when comparing the blueshifted and redshifted velocity intervals, the overall spatial distribution of CO($J$=6--5) is broadly similar to that of the other tracers and is markedly asymmetric: rather than being distributed quasi-isotropically around the protostar, most of the emission is concentrated toward the northeast. This morphology differs substantially from that of cold and dense gas tracers in the same region, such as N$_2$D$^{+}$ and H$_2$D$^{+}$, which are elongated approximately along the north--south direction \citep{Tokuda_2020Tau,Tokuda_2025}. The densest component detected in our new Band~9 data, i.e., the continuum emission shown in Figure~\ref{fig:COHCOP}(e), lies at a higher density than the gas traced by CO($J$=6--5). While the noise level decreases toward the edges of the field due to the primary-beam response, the robust emission is confined to the vicinity of the protostar with $\gtrsim$3$\sigma$(= 11\,mJy\,beam$^{-1}$). We interpret this Band~9 continuum source as the blended emission from the previously identified protostellar source MMS--1 and the nearby starless condensation MMS--2. The emission is dominated by MMS--2 and appears spatially connected to the CO($J$=6--5) ring. The fainter condensation MMS--3 is not detected in the Band~9 continuum.

\begin{figure}[htbp]
    \centering
    \includegraphics[width=1.0\columnwidth]{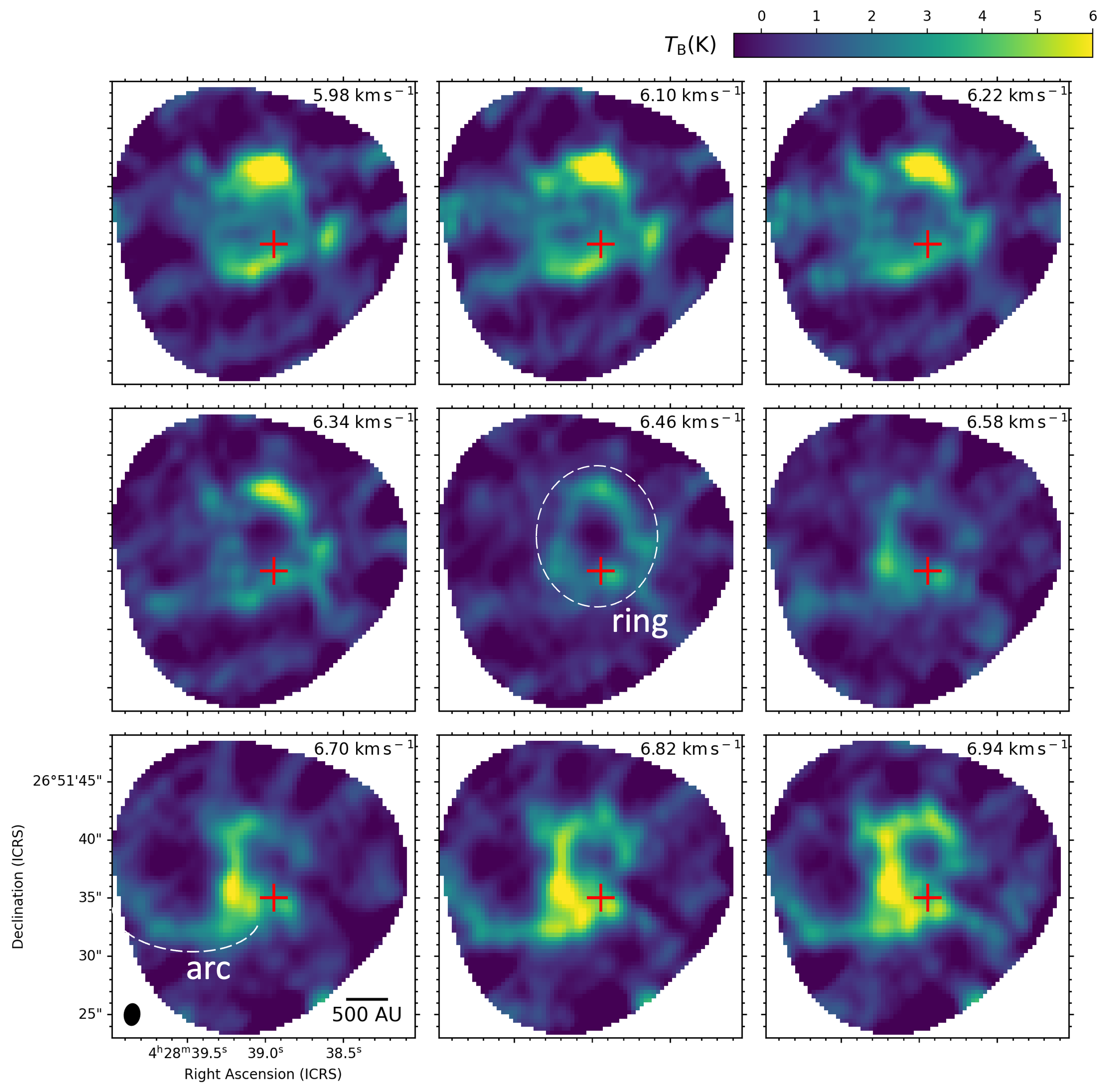}
    \caption{
    Channel maps of the CO($J$=6--5) emission at the native spectral resolution ($\Delta v=0.12~{\rm km~s^{-1}}$), shown as a $3\times3$ panel. All panels are displayed in brightness temperature units (K). The central velocity channel is shown in the upper right corners of each panel in km\,s$^{-1}$ unit. The beam size of $1\farcs8 \times 1\farcs3$ (P.A.$=-6\fdg1$) is shown in the ellipse at the lower left corner in the lower left panel. The red cross shows the position of the protostar.}
    \label{fig:CO65chanmap02}
\end{figure}

Figure~\ref{fig:CO65chanmap02} presents channel maps around the systemic velocity at the native spectral resolution of $\Delta v=0.12~{\rm km~s^{-1}}$. In these channels, the ring-like feature persists continuously from $\sim$6 to 7~km~s$^{-1}$, suggesting that it is a coherent component in space and velocity. 
As a representative example, we highlight the ring-like morphology in the $V_{\rm LSRK}=6.46~{\rm km~s^{-1}}$ channel with a dotted ellipse. 
Higher-density tracers such as H$^{13}$CO$^{+}$($J$=3--2)\citep{Tokuda_2014} show some spatial and velocity overlap with the CO($J$=6--5) emission on the side closer to the protostar, but they do not reproduce the closed ring-like morphology. This selectivity suggests that the ring emission arises from a relatively restricted range of physical conditions, requiring elevated temperature and/or density compared to the bulk of the cold dense core (see more details \ref{subsec:ring_phys}). An arc-like feature is evident toward the east at $V_{\rm LSRK}\simeq 6.7$--$7.0~{\rm km~s^{-1}}$. 

\begin{figure}[htbp]
    \centering
    \includegraphics[width=1.0\columnwidth]{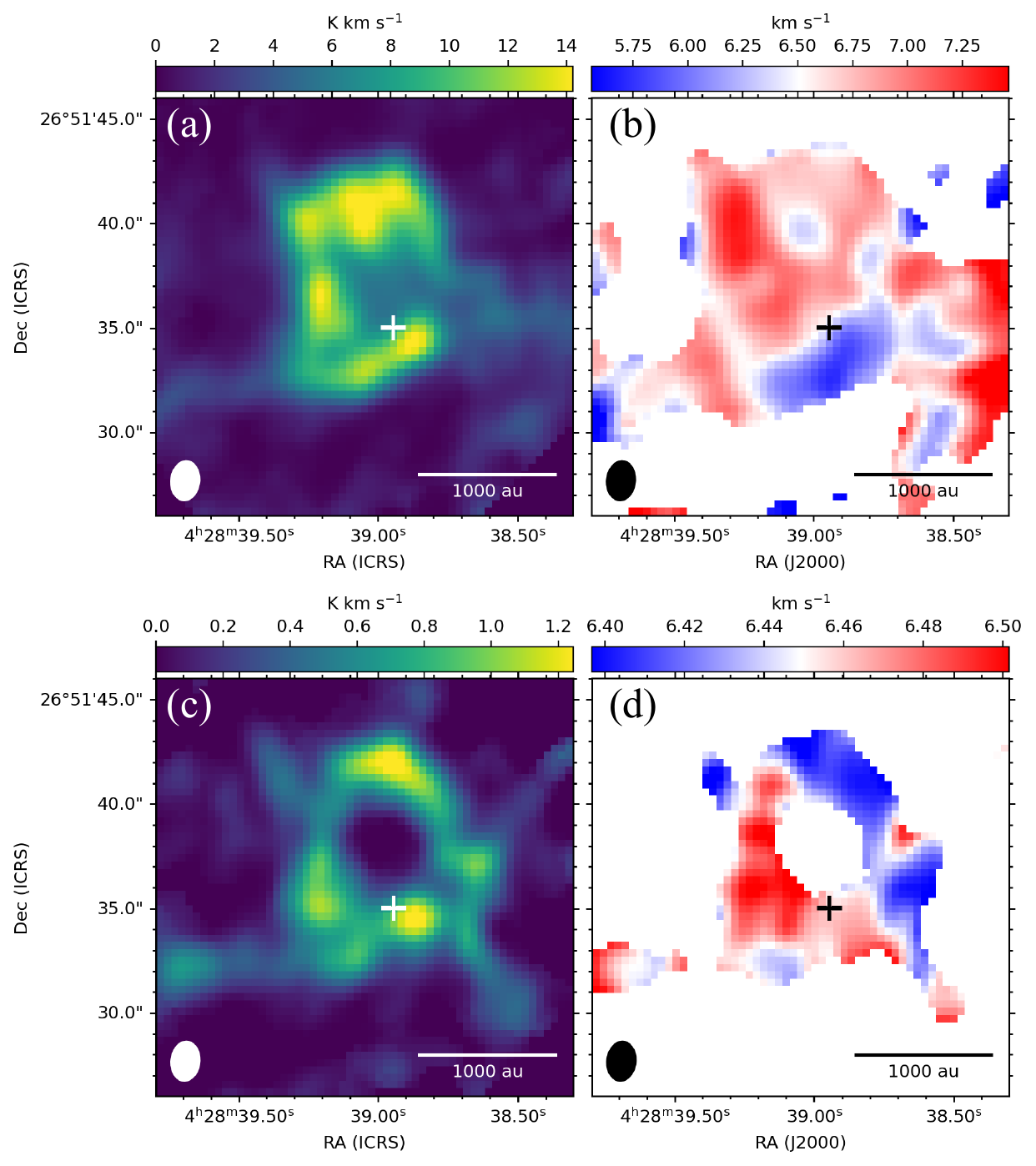}
    \caption{{\bf (a)} Integrated-intensity (moment~0) map of CO($J$=6--5) integrated over $V_{\rm LSRK}=4.4$--$8.6~{\rm km~s^{-1}}$. The beam size ellipse is shown in the lower left corner. The cross mark shows the protostar position. 
    {\bf (b)} Intensity-weighted mean velocity (moment~1) map computed over the same velocity range. 
    {\bf (c)} Same as panel (a), but for the integrated velocity range of $V_{\rm LSRK}=6.3$--$6.7~{\rm km~s^{-1}}$.
    {\bf (d)} Same as panel (b), but the velocity range as same as in panel (c).}
    \label{fig:CO65mom_maps}
\end{figure}

To quantify the morphology and velocity structure of the ring and its surrounding gas, we visualize the emission in two complementary velocity ranges (Figure~\ref{fig:CO65mom_maps}). Figures~\ref{fig:CO65mom_maps}(a) and (b) present the broader velocity view. The ~0 moment map (panel~a) outlines a limb-brightened structure. A faint but significant emission is also detected toward its interior. The corresponding moment~1 map (panel~b) indicates that gas immediately south of the protostar is blueshifted to $V_{\rm LSRK}\sim5~{\rm km~s^{-1}}$, while a more redshifted component ($\gtrsim$7\,km\,s$^{-1}$) is present toward the north.

Figures~\ref{fig:CO65mom_maps}(c) and (d) focus on velocities closer to the systemic velocity, highlighting the channels where the ring morphology is most clearly defined. As shown by the moment~0 map integrated over $V_{\rm LSRK}=6.3$--$6.7~{\rm km~s^{-1}}$ (panel~c), the ring becomes significantly sharper, with a central cavity of diameter $\sim$1000\,au. The typical peak brightness temperature at these velocity components is $\sim$3~K. The moment~1 field (panel~d) also exhibits a distinct pattern compared to panel~(b): the eastern side of the ring is systematically redshifted, whereas the western side is blueshifted. We discuss the implications of this velocity structure in Section~\ref{subsec:ring_kin}.

\section{Discussion} \label{sec:D}

Our Band~9 CO($J$=6--5) ACA data reveal a $\sim$1000~au-scale ring-like structure located off-center from the protostar and appearing most clearly at velocities close to the systemic velocity. To our knowledge, such an off-centered ring in a high-$J$ CO transition has not been previously reported in the context of the earliest protostellar phase among Class~0 objects. In this section, we discuss plausible physical properties and kinematics of the ring, and consider how this structure may fit into the earliest stages of protostellar evolution.

\subsection{Physical properties of the ring near the systemic velocity} \label{subsec:ring_phys}

The CO($J$=6--5) ring appears spatially and kinematically connected to the dense-gas component traced by H$^{13}$CO$^{+}$($J$=3--2) \citep{Tokuda_2014}. Motivated by this association, we infer that the CO($J$=6--5) ring traces gas denser than the ambient cold envelope but less dense than the compact condensations,  MMS--2 and MMS--3. In particular, MMS--2 located southwest of the protostar has been suggested to reach $n({\rm H_2})\sim10^{7}~{\rm cm^{-3}}$, with its surroundings at $n({\rm H_2})\sim10^{6}~{\rm cm^{-3}}$ \citep{Tokuda_2014,Tokuda_2016}. Adopting these values as guides, we infer that the characteristic density of the CO($J$=6--5) ring is likely $n({\rm H_2})\sim10^{5}$--$10^{6}~{\rm cm^{-3}}$, comparable to the critical density of CO($J$=6--5) at $T_{\rm kin}\sim10$--100~K \citep{Goorvitch_1994}.

The observed CO($J$=6--5) emission has a typical peak brightness temperature of a few kelvin, which provides a useful constraint on the temperature of the emitting gas. For illustration, if the gas were as cold as $T_{\rm kin}=10$~K and fully thermalized (i.e., under LTE with optically thick emission), the maximum line brightness would be limited to the Rayleigh--Jeans-equivalent temperature of a 10~K blackbody at this frequency, $T_{\rm B}\simeq1.2$~K. Therefore, if the CO($J$=6--5) emission arises from gas near 10~K, even the brightest possible emission would fall well below the observed values. Explaining the observed $T_{\rm B}\sim$ a few kelvin thus requires higher temperatures, plausibly $T_{\rm kin}\gtrsim20$~K. At the densities inferred here ($n({\rm H_2})\sim10^{5}$--$10^{6}\,{\rm cm^{-3}}$), gas-phase CO is expected to become significantly depleted onto dust grains at low temperatures \citep[e.g.,][]{Bergin_2007}. Therefore, if the ring component were colder than $\sim$20\,K, CO freeze-out would likely reduce the observable CO emission, providing additional support for $T_{\rm kin}\gtrsim$20\,K. This inference is consistent with the presence of high-brightness temperature ($\sim$30--70~K) components traced by CO($J$=3--2) at velocities offset from the systemic velocity \citep{Tokuda_2018}. The bottom panel of Figure~\ref{fig:spect} shows the CO($J$=3--2) Total Power spectrum, which exhibits only $\sim$2~K around $V_{\rm LSRK}\sim7~{\rm km~s^{-1}}$. This low apparent brightness likely reflects the dominance of an optically thick, low-excitation component and strong self-absorption near the systemic velocity in the low-$J$ transition.

In summary, the current constraints favor $n({\rm H_2})\sim10^{5}$--$10^{6}~{\rm cm^{-3}}$ and $T_{\rm kin}\gtrsim20$~K for the systemic-velocity ring-like component and its immediate surroundings. Because the ring lies at velocities where low-$J$ CO interferometric data are strongly affected by optical depth and self-absorption, direct multi-transition line ratios are not yet available to tighten physical constraints. Follow-up observations at higher angular resolution with the 12~m array and additional high-$J$ transitions (e.g., CO $J$=7--6 in Band~10) would substantially improve the determination of the density and temperature of this component.

\subsection{Velocity structure of the ring and its kinematic implications} \label{subsec:ring_kin}

\begin{figure}[htbp]
    \centering
    \includegraphics[width=1\columnwidth]{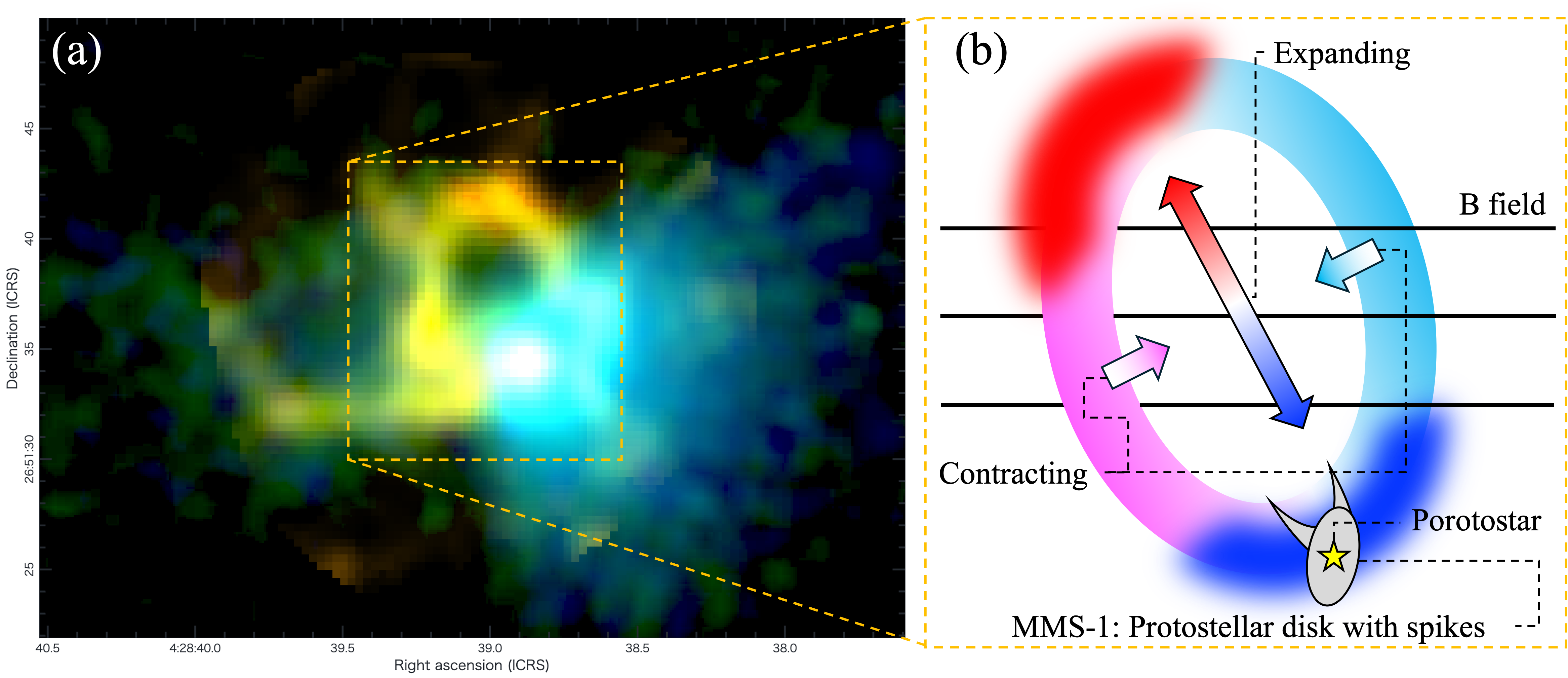}
    \caption{
    {\bf (a)} The background image is constructed from Spitzer bands with $4.5~\mu$m shown in green and $3.6~\mu$m shown in blue \citep{Bourke06}. Orange color shows the integrated intensity (moment~0) of the CO($J$=6--5) emission integrated over $V_{\rm LSRK}=6.2$--$6.9~{\rm km~s^{-1}}$.
    {\bf (b)} The colored ring indicates the qualitative line-of-sight velocity pattern, with red/magenta and blue/cyan denoting redshifted and blueshifted emission, respectively. The star marks the position of the protostar, which is associated with a spike-like protostellar disk structure \citep{Tokuda_2024}. The arrows highlight the characteristic sense of motion discussed in Section~\ref{subsec:ring_kin}. The black solid lines indicate the adopted orientation of the magnetic-field direction on the plane of the sky, guided by single-dish dust polarization measurements \citep{Fukaya23}.}
    \label{fig:ponti}
\end{figure}

Here we place the CO($J$=6--5) ring in the broader multi-wavelength context of the system. As shown in Figure~\ref{fig:ponti}(a), the ring is located between the bipolar nebula seen in the Spitzer image. The scattered-light emission in the shorter wavelength (3.6\,$\mu$m) is brighter on the western side, suggesting that the western outflow cavity is the near side, while the eastern cavity is the far side \citep{Bourke06,Terebey09}. The location of the CO($J$=6--5) ring is also aligned with the direction of the spike-like structure extending from the protostellar disk, MMS-1 \citep{Tokuda_2024}. Motivated by these geometric considerations, we adopt the schematic configuration illustrated in panel~(b): the ring is expanding along its projected major axis while undergoing a slight contraction along its projected minor axis. In what follows, we describe the observational reasoning that led us to this interpretation.

Ring-like gas morphologies in protostellar systems have, in some cases, been interpreted as outflow-related structures viewed close to face-on \citep[e.g.,][]{Harada_2023}. One might therefore ask whether the CO($J$=6--5) ring in MC~27 could reflect a similar geometry. However, this interpretation is disfavored for MC~27. First, the CO($J$=6--5) ring is dominated by systemic-velocity emission and shows no clear connection to the outflow-wing component. Second, the compact dust disk is highly inclined ($i\sim70^\circ$; \citealt{Terebey09,Tokuda_2017,Tokuda_2024}), disfavoring a nearly face-on outflow geometry.

These eliminative arguments motivate us to revisit the scenario proposed by \citet{Tokuda_2024}, in which magnetic flux and gas are advected outward from the disk/envelope interface through interchange instability \citep{Parker_1979,Kaising_1992,Lubow_1995,Stehle_2001}. In this scenario, strong-field regions appear around the outer edge of the disk due to the redistribution of the magnetic field by non-ideal magnetohydrodynamic (MHD) effects; when the magnetic pressure locally exceeds the ram pressure of the accreting gas, the interchange instability develops and expels magnetic flux from the disk (or high-density region) into the envelope (or the low-density region). The expelled magnetic flux drives the formation of a ring-like structure that expands at roughly the sound speed \citep{Zhao_2011,Matsumoto_2017}, and its densest portion appears observationally as the spike-like feature associated with the disk (Figure~\ref{fig:ponti}(b)). In this framework, the magnetic field retained within the evacuated interior is expected to be approximately perpendicular to the ring plane. Single-dish dust polarization observations of this system indeed indicate a plane-of-sky magnetic-field orientation in the immediate vicinity of the protostar that is aligned roughly east--west \citep{Fukaya23}. This observed field geometry is consistent with, and lends additional support to, the presence of a magnetic field threading the ring.

Moreover, if the interchange instability occurs recurrently and releases magnetic flux episodically, the ring's interior may remain magnetically dominated. In that case, the gas velocity should be comparable to Alfv\'en speed and exceed the thermal sound speed, potentially accounting for the faster apparent expansion inferred along the major axis. For example, in the interchange-driven system modeled by \citet{Machida_2025}, the magnetic field strength at radii of a few hundred au can reach $B\sim$10$^{-3}$\,G (see their Figure~5). If the lower-density interior of the ring has $n({\rm H_2})\lesssim10^{5}~{\rm cm^{-3}}$, the corresponding Alfv\'en speed is given by the following equation, 
\begin{equation}
v_A \simeq 4.5~{\rm km~s^{-1}}
\left(\frac{B}{10^{-3}~{\rm G}}\right)
\left(\frac{n({\rm H_2})}{10^{5}~{\rm cm^{-3}}}\right)^{-1/2}.
\end{equation}
For a simple geometric correction, we parameterize the tilt of the expansion direction relative to the plane of the sky by an angle $\phi$ (with $\phi=0^\circ$ for motion in the sky plane and $\phi=90^\circ$ for purely line-of-sight motion). The observed maximum velocity offset is $\sim$3\,km\,s$^{-1}$ (Figures~\ref{fig:spect}, and \ref{fig:COHCOP}). Assuming that the ring is not strongly tilted relative to the plane of the sky ($\phi\le45^\circ$), we obtain $v_{\rm exp}\gtrsim$4.2\,km\,s$^{-1}$, broadly consistent with the expected Alfv\'en speed for the inferred magnetic-field strength and gas density. In addition, \citet{Tokuda_2018} argued, using the Rankine--Hugoniot jump conditions, that shock heating under comparable physical conditions can reproduce the elevated gas temperatures, $\sim$30--70~K, inferred observationally. Indeed, toward the location of the prominent redshifted blob, shock-related tracers such as CH$_3$OH and SO have also been detected \citep{Favre_2020}, further supporting an interpretation involving localized shocks.

If we assume that the disk orientation inferred from the scattered-light outflow cavities is approximately consistent with the inclination of the CO($J$=6--5) ring, then the western side of the ring corresponds to the far side and the eastern side to the near side. Under this geometry, the observed line-of-sight velocity pattern suggests that the ring is undergoing a slight contraction (Figure~\ref{fig:ponti}). Such behavior is naturally expected if the ring is confined by the surrounding dense material. Indeed, theoretical calculations also show that once a ring-like structure forms, its subsequent evolution can be anisotropic: depending on the direction, parts of the ring can undergo contraction or exhibit oscillatory motions rather than continuing a purely monotonic expansion \citep[e.g.,][]{Machida_2025}.
Note that the arc-like structure seen near $V_{\rm LSRK}\sim$6.7\,km\,s$^{-1}$ is likely a remnant of an earlier interchange-instability event, as suggested by \citet{Tokuda_2024}; although it may have grown into a larger-scale structure, the gas becomes more diffuse at larger distances from the protostar, such that only a portion of the shell is visible as an arc rather than a closed ring.

\subsection{Broader implications of detecting a warm ring in the context of star formation and comparisons with other sources}
\label{subsec:ring_mean}

Candidates for ring- or bubble-like structures that may be produced by drastic magnetic-flux redistribution driven by interchange instability are increasingly being reported in both observations and theory \citep{Matsumoto_2017,Machida_2020,Machida_2025,
Tokuda_23interC,Tanious_2024,Fielder_2024}. One of the most striking examples of observations is the CrA IRS~2, where a remarkably clear ring is detected in C$^{18}$O($J$=2--1) with a diameter of $\sim$7000~au \citep{Tokuda_23interC}. Because IRS~2 is a Class~I object, it represents a more evolved stage than MC~27 and likely traces comparatively moderate-density ($\sim$10$^{4}$\,cm$^{-3}$) and cooled ($\sim$10\,K) material in its extended environment. High-angular-resolution dust-continuum imaging of the IRS~2 disk, with a radius of $\sim$60\,au, has also revealed ring--gap substructures \citep{Shoshi_2026}, a feature not commonly reported in Class~I systems \citep[e.g.,][]{Ohashi23}. \cite{Shoshi_2026} argues that enhanced magnetic dissipation in the disk may suppress the magnetorotational instability and thereby facilitate early planet formation. In contrast, the protostellar disk inferred in MC~27 is extremely compact ($\sim$10~au), and the strong magnetic field accumulated around the system could instead imply efficient magnetic braking that keeps the disk small \citep{Tokuda_2024}. Taken together, these results suggest that drastic magnetic-flux redistribution may offer new clues not only to the emergence of ring- or bubble-like structures in the embedded phase but also to the subsequent evolution of disks, including their sizes and substructures that set the initial conditions for planet formation.

A natural question is how common such ring structures may be. At present, it is premature to discuss their universality because the physical regime suggested for the CO($J$=6--5) ring in MC~27, namely gas with $n({\rm H_2})\sim10^{5}~{\rm cm^{-3}}$ and $T\gtrsim20$~K, is difficult to probe systematically: Band~9 (and/or further higher-frequency) observations remain observationally demanding, and spatially resolved imaging of high-$J$ CO in embedded protostars is still sparse. Nevertheless, hints may be available from other young sources in nearby star-forming regions. For example, IRAS~15398-3359 has an evolutionary stage comparable to the protostar in MC~27 and may even be younger when judging its protostellar mass \citep{Okoda_2018}. A ring-like structure has been reported around this source in H$^{13}$CO$^{+}$ \citep{Jorgensen_2013}. 
While an accretion-burst interpretation has been discussed in the literature, the reported ring center appears slightly offset from the protostar. This geometry does not uniquely point to a single origin and leaves room for alternative interpretations, including magnetically driven redistribution of flux and gas. Similarly, \citet{Mercimek_2023} noted that in VLA1623-2417 a ring-like feature appears to connect components A/B and W (see their Figure~4), suggesting that ring morphologies may arise in dynamically complex, magnetized multiple systems as well.

More broadly, although low-$J$ CO and its isotopologues have greatly advanced our understanding of Class 0/I envelope structures, interpreting the systemic-velocity gas in low-$J$ CO is often difficult. High-frequency CO observations provide a complementary and potentially crucial perspective on the embedded protostellar stage, offering insight into the frequency of warm rings/arcs and related magnetically regulated structures in the earliest phases of star formation.

\section{Summary}\label{sec:sum}

We observed the cold protostellar core MC~27/L1521F in Taurus, which hosts a Class~0 protostar, in the CO($J$=6--5) line using the ALMA ACA 7~m array with the Band~9 receiver. We detect, for the first time, a ring-like structure at velocities close to the systemic velocity. Together with the previously known arc feature in this system, the ring-like emission forms a visually distinctive pattern with the Arabic numeral ``9''. From multiple lines of evidence and excitation calculations, we infer that the ring traces relatively dense and warm gas, with $n({\rm H_2})\sim10^{5}$--$10^{6}~{\rm cm^{-3}}$ and $T_{\rm kin}\gtrsim20$~K. Forming an off-centered ring on $\sim$1000~au scales is difficult to reconcile with standard outflow-driven scenarios alone. Instead, our results favor an interpretation in which magnetic-flux redistribution from the disk/envelope interface, potentially driven by interchange instability, generates expanding structures and associated shock heating.

\begin{acknowledgments}

We would like to thank the anonymous referee for useful comments that improved the manuscript. We are grateful to Shu-ichiro Inutsuka, Tsuyoshi Inoue, Kengo Tomida, Takashi Hosokawa, Hiroko Shinnaga, and Shingo Nozaki for helpful discussions and insightful comments on the observational characteristics and theoretical interpretation of MC~27/L1521F. This paper makes use of the following ALMA data: ADS/JAO, ALMA\#2011.0.00611.S, 2015.1.00340.S and 2024.1.00023.S. ALMA is a partnership of ESO (representing its member states), the NSF (USA), and NINS
(Japan), together with the NRC (Canada), MOST, and ASIAA (Taiwan), and KASI (Republic of Korea), in cooperation with the Republic of Chile. The Joint ALMA Observatory is operated by the ESO, AUI/NRAO, and NAOJ. This work was supported by a NAOJ ALMA Scientific Research grant Nos. 2022-22B, Grants-in-Aid for Scientific Research (KAKENHI) of Japan Society for the Promotion of Science (JSPS; grant Nos. JP18H05440, JP20H05645, JP21H00049, JP21K13962, JP23H00129, JP25KJ1921, and JP25K07369), and Kagawa University
Research Promotion Program 2025 grant No. 25K0D015. We acknowledge the use of OpenAI's ChatGPT only as a grammar-checking and editing tool to improve the clarity and readability of the manuscript.

\end{acknowledgments}

\software{astropy \citep{Astropy22}, CASA \citep{CASAteam_2022} }

%\appendix

\end{document}